\documentstyle[psfig]{article}

\begin{document}

\begin{center}
\begin{large}
APERTURE PHOTOMETRY OF SATURATED STAR IMAGES FROM DIGITISED PHOTOGRAPHIC PLATES
\end{large}

\vspace{.7cm}

{\it By J.L. Innis\\
280 Brightwater Rd, Howden, Tasmania, 7054, Australia \vspace{0.2 cm} \\
D.W. Coates \\
School of Physics and Materials Engineering, Monash University, Clayton, Victoria, 3800, Australia\vspace{0.2 cm}\\
A.P. Borisova and M.K. Tsvetkov\\
Institute of Astronomy, 72 Tsarigradsko Shosse Blvd., 1784 Sofia, Bulgaria\\}

\end{center}

\vspace{0.35 cm}

\begin{large}
Saturated stellar images on digitised photographic plates are many times greater in area than the  `seeing disk' seen in unsaturated CCD images. Indeed the flux profile of a bright star can be traced out for several degrees from the star's centre$^{\rm {1,2,3}}$.  The radius of the saturated stellar image can often be directly related to the magnitude of the star, a fact well known and exploited in iris photometry$^{\rm {4,5}}$.  In this work we compare the radial flux profile of stars in the approximate range B$\sim$9 to $\sim$13 mag, obtained from scans of plates from the Bamberg Sky Patrol archive, with a profile of the form measured by King$^{\rm {1}}$.  We show that simple aperture photometry of saturated stellar images, obtained from photo-positives of scanned photographic plates, yield data that are in agreement with simulations using a (saturated) synthetic stellar radius profile.  Raw plate magnitudes from this aperture photometry can be easily and satisfactorily transformed to standard magnitudes, as demonstrated in a recent study carried out by the current authors$^{\rm {6}}$.  
\end{large}

\vspace{0.35 cm}
{\it Introduction}\\

\begin{figure} 
\psfig{file=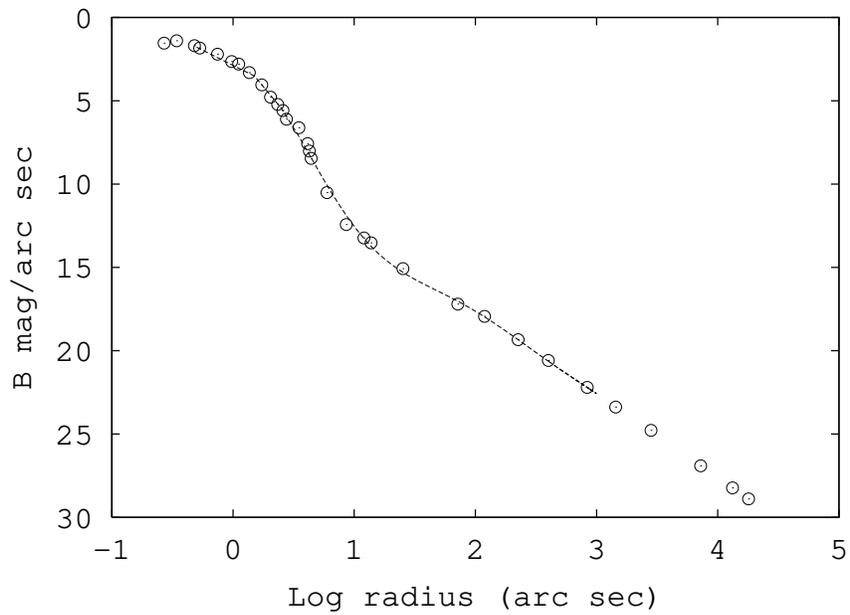,width=12cm,angle=-90}
\caption{ Stellar profile, in {\it B} magnitudes per square arc second, from 
King$^{\rm {1}}$. Open circles represent data points digitised from King's Figure 
1. The line represents a polynomial fit to the data (from 0.5 arc second 
radius outwards) used to create the synthetic stellar profiles employed in the 
current work.}
\label{King}
\end{figure}

The radial profile of a star observed by a ground-based imaging system is known to be far more extended than would be expected from the instantaneous seeing disk or from diffraction effects in the instrument.  King$^{\rm {1}}$, Kormendy$^{\rm {2}}$ and Racine$^{\rm {3}}$ provided examples of the stellar profile, and discussed possible causes. It appears that both instrumental and atmospheric effects are likely to be contributing factors$^{\rm {3}}$. Fig.~\ref{King} shows the profile derived by King (open circles) for a zeroth--magnitude star from photographic data, together with a polynomial fit to the data (solid line) which we will use below as our synthetic profile.  We have adopted this, rather than the more recent, CCD-derived, profile of Racine$^{\rm {3}}$ as we are concerned with aperture photometry of scanned photographic plates.

Significantly, for radii greater than about 10 arc seconds, there is an approximate power law relation between flux (in magnitude per arc second) and radius$^{\rm {7}}$.  Hence, for iris photometry of a saturated stellar image of radius of 10 arc sec or more, when the star is centred in the iris diaphragm, there is a linear relationship between the logarithm of the measured transmission and magnitude$^{\rm {5}}$. There is a direct, but inverse, relationship between the darkening of the original photo--negative plate and the brightening of a photo--positive image. Consequently we might expect, for a digitised {\it photo--positive} plate scan, a near--linear relationship between the plate magnitude (logarithm of the `counts' from aperture photometry) and stellar magnitudes in this case.  

\vspace{0.35 cm}
{\it Aperture photometry of photo-positive plate scans}\\

We explored the above suggestion using a series of scans made of Bamberg Sky Patrol plates ({\it e.g.} see Strohmeier and Mauder$^{\rm {8}}$), from the plate archive of the Dr. Remeis Observatory, Bamberg.  The scans were originally made for a study of the active star CF Octantis.   A short description of the Sky Patrol programme and more details of the CF Octantis results are presented in Innis et al.$^{\rm {6}}$. In brief, the Sky Patrol plates were obtained with 0.1-m diameter cameras at three Southern Hemisphere sites between 1962 and 1976, and have a typical limiting photographic magnitude near 14. An Epson flat-bed scanner was used to digitise 375~plates containing the field of CF Oct, at a resolution of 5.25 arc sec per pixel. The scans were converted to photo-positives and saved as FITS files. The analysis package IRAF was used to obtain aperture photometry of 9 field stars (ranging from B$\sim$9 to 10.8) and CF Oct on 353 plate scans, with the other 22 plates being unusable due to cloud, plate scratches, or other effects.

\begin{figure} 
\psfig{file=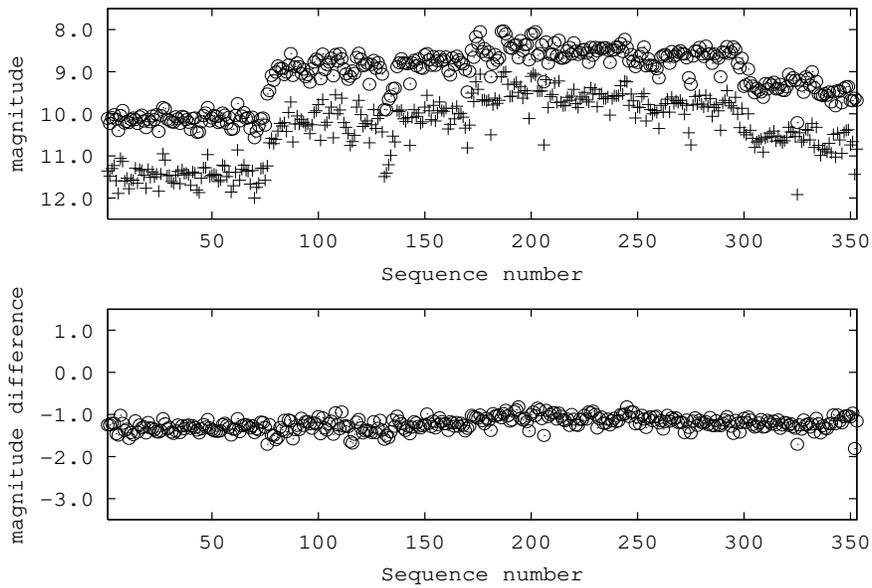,width=12cm,height=8cm,angle=0}
\caption{Example sequence of raw, untransformed, plate magnitudes, from IRAF aperture 
photometry, of two stars, measured from $\sim$350 digitised Bamberg sky patrol 
plates from 1965 to 1969. Both stars showed saturated cores on the plates. Top panel: raw plate magnitudes of HD 195460 (circles) and CP $-$80~980 ($+$ symbols). The large jumps near sequence 
numbers 75 and 300 correspond to annual breaks in the plate runs. Lower panel: 
Differential magnitudes between these two stars, plotted on the same vertical scale as the upper panel. The standard deviation of the differences was 0.16~mag. We measured up to 19 stars 
on each plate, but for clarity results for two stars only are plotted here. 
Although the zero points change significantly from plate to plate, by nearly 2 
magnitudes, the magnitude differences remained much more consistent.}
\label{mag_time_series}
\end{figure}

Example raw plate magnitudes for two field stars, HD 195460 (B$\sim$8.99) and CP $-$80~980 (B$\sim$10.62) are shown in the top panel of Fig.~\ref{mag_time_series}.  Both stars showed saturated cores on the plates. We used a fixed aperture of 20 pixel radius ($\sim$105~arc seconds) for the measurements (see Innis et al.$^{\rm {6}}$ for more details). Large and sudden vertical shifts in the mean levels correspond to annual breaks in the data runs. Data from three different sites are shown.  While the magnitudes vary significantly from plate to plate, by up to $\sim$2 magnitudes, they do so in a reasonably consistent manner.  The magnitude differences (lower panel, same vertical scale) are much more uniform. For these two stars the standard deviation of the differences was 0.16~mag.  As noted, 9 field stars (and CF Oct) were measured on each plate. For clarity of presentation results for only two field stars are shown, but these illustrate typical behaviour.  As part of a further study, 10 additional field stars (down to B$\sim$13) were measured on a subset of 90 plates.  Similar results were found, albeit with slightly increased observational errors.  Combining the two field star data sets for the 90 plates gives plate magnitudes for 19 stars ranging from B$\sim$9 to $\sim$13, which forms the basis of our present study.

\newpage
\vspace{0.35 cm}
{\it Synthetic stellar profiles} \\

\begin{figure} 
\psfig{file=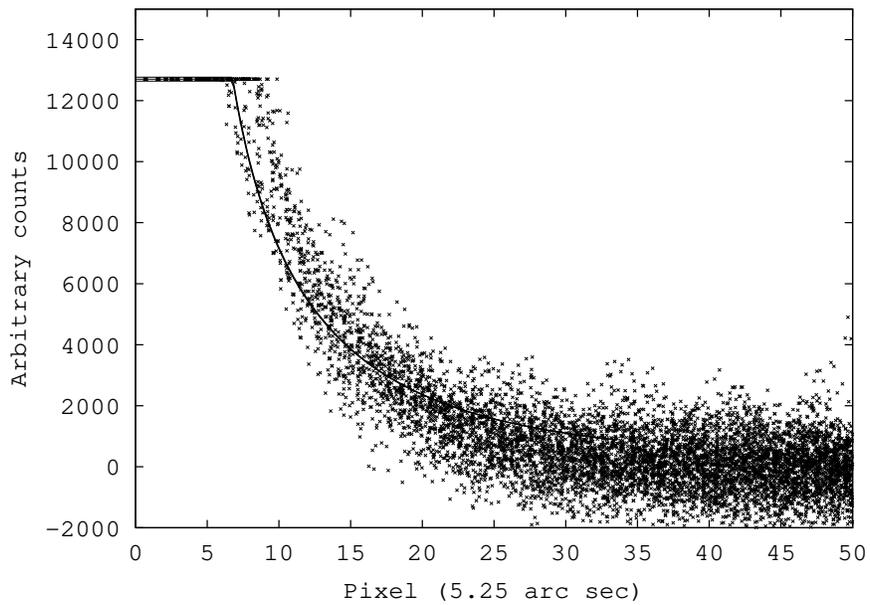,width=12cm,angle=0}
\caption{Radial profile plot showing a comparison between an observed 
(sky-subtracted) saturated stellar profile from a digitised Bamberg plate (points) 
and a synthetic profile (solid line) derived from the polynomial fit to the data of King 
(1971).  The synthetic profile is not a fit to the Bamberg data. The saturation limit of the synthetic profile was chosen so that the radius of saturation (in units of pixels of 5.25 arc seconds square) approximately matched the observed radius of saturation.}
\label{profile_comparison}
\end{figure}

Fig.~\ref{profile_comparison} shows a radial plot of an observed (`sky subtracted') stellar profile (square symbols), obtained from a digitised plate, plotted as arbitrary intensity versus 5.25 arc second pixels.  The observed profile is saturated out to about 6 or 7~pixels in radius. Also plotted is part of the polynomial fit to the King profile (from Fig.~\ref{King}), with a saturation limit imposed so that the radius of saturation is 6~pixels to approximately match the observed profile, and then scaled vertically so that the saturation intensities match.  The agreement is reasonable although the observed profile falls more rapidly than the synthetic profile at larger radii. The difference at large radius is, at least in part, probably a consequence of our polynomial fit (see Fig.~\ref{King}), which appears slightly high near 100 arc sec ($\sim$20~pixels). For the purposes of this study we believe the small difference can be neglected, as we are attempting to illustrate a general principle rather than derive an exact match to the stellar profiles recorded on the Bamberg plates.

\begin{figure} 
\psfig{file=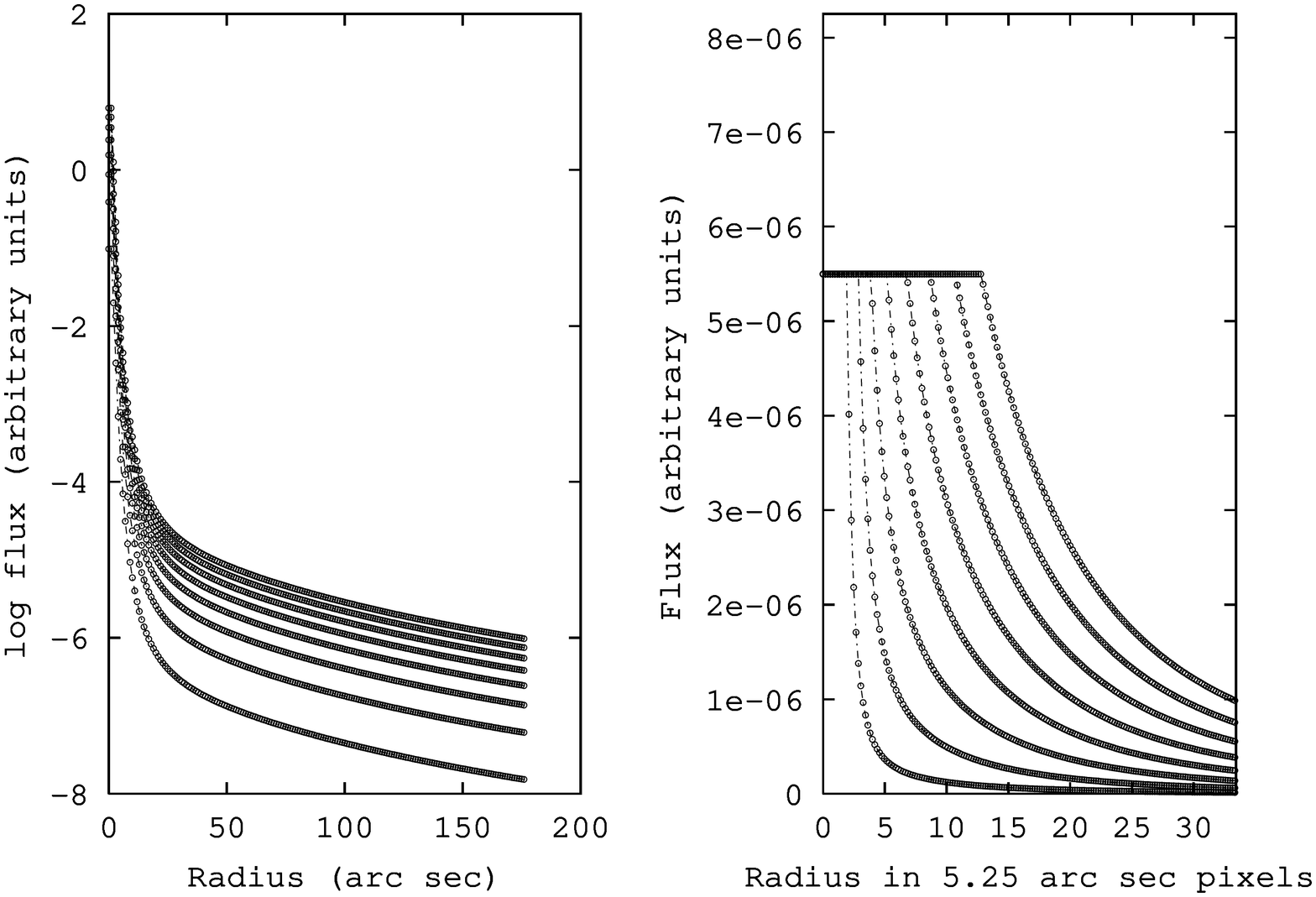,width=12cm,height=7cm,angle=-90}
\caption{ Synthetic stellar flux profiles, derived from the fit to the data of King 
(1971).  Left panel -- a series of profiles spanning $\sim$4~magnitudes, 
plotted on a semi--log scale. Flux is in arbitrary units, while the horizontal 
axis is in arc seconds.  Right panel -- the same profiles but with a common 
`saturation limit' imposed. The horizontal scale in this panel is now in units 
of 5.25 arc second square pixels to match the resolution of the scanned 
Bamberg plates.}
\label{synthetic}
\end{figure}

We generated a series of synthetic stellar radial--flux profiles, based on our fit to the data of King$^{\rm {1}}$, spanning a range of 4 magnitudes.  These are shown in the left panel of Fig.~\ref{synthetic} on a semi-log plot, with the x--axis in arc seconds.  We then applied a common `saturation limit' to all profiles to mimic the saturation level of the photographic emulsion, and resampled the data to a 5.25 arc second pixel size. The right hand panel of Fig.~\ref{synthetic} shows the resulting profiles.  The radius of saturation varies from around 2 to 13 pixels.  As explained below, the saturation intensity level was chosen so that the resulting radii of saturation matched the observed values from the plate scans.  We stress that the only adjustable parameter in this matching is the choice of saturation intensity level, and this is applied to all the synthetic profiles.

\vspace{0.35 cm}
{\it Radii of saturation}\\

\begin{figure} 
\psfig{file=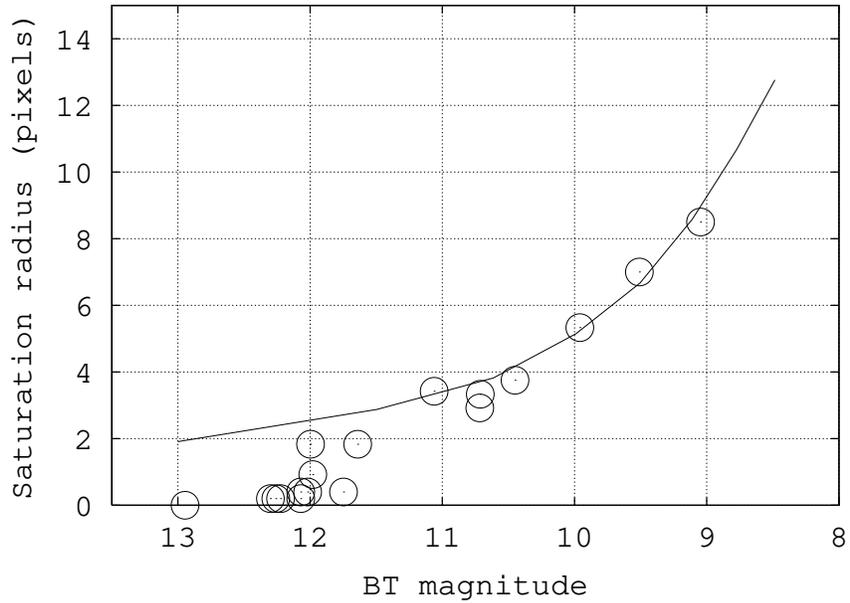,width=12cm,angle=-90}
\caption{A comparison between the observed `radius of saturation' (open circles) 
for 19 stars ranging from Tycho--2 BT magnitudes of $\sim$13 to $\sim$9, and the calculated radius of saturation (solid line) found using the King profiles.  The flux level where 
saturation occurred was the only parameter adjusted to give the calculated 
curve, and was chosen on the basis of the comparison shown in  
figure~\ref{profile_comparison}. The observed saturation radii are based on 
averages from 6 `average' plates.  The disparity at the lower radii values may 
in part be a consequence of the limited resolution of the plate scan (5.25 arc 
second pixels), while the calculated radii were derived from a 1 arc second grid and then rebinned.}
\label{radius_comp}
\end{figure}

Fig.~\ref{radius_comp} shows, as open circles, the measured radii of saturation for our 19 field stars, ranging from B$\sim$9 to $\sim$13.  We measured radii on 6 digitised plates whose plate magnitudes were close to the ensemble mean determined from 90 plates, and then took averages for each star.  This procedure should minimise possible changes in stellar profiles caused by
instrumental changes made during the survey. The calculated radii of saturation of the synthetic profiles of Fig.~\ref{synthetic} are represented by the solid line. We adjusted the common saturation level for the synthetic profiles to get a reasonable match (by eye) between the calculated and measured radii of saturation. As noted, this is the only parameter that we adjusted, after fixing a zero--point for the synthetic magnitudes. The agreement is reasonable for the larger radii, but is poor for the fainter stars. In part this may be a consequence of the limited resolution of the plate scans: The synthetic profiles were calculated on a 1--arc second axis, then resampled to a 5.25 arc second pixel scale to match the plate scan resolution.  When the area of saturation of the profile on the plate is comparable to the 5.25$\times$5.25 arc sec resolution of the scan, undersampling starts to reduce the observed saturation radius, hence Fig.~\ref{radius_comp} shows poorer agreement for the fainter stars.

\newpage
\vspace{0.35 cm}
{\it Comparison of synthetic and aperture photometry}\\

\begin{figure} 
\psfig{file=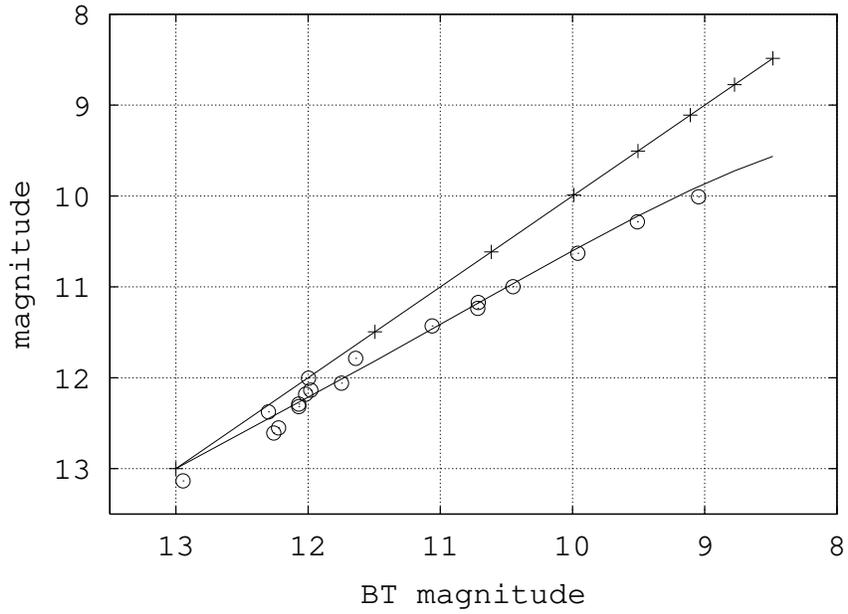,width=12cm,angle=-90}
\caption{Mean measured plate magnitudes for 19 stars (circles), based on aperture photometry with IRAF, from 90 digitised Bamberg plates, and calculated magnitudes from aperture photometry (solid line, no symbols) of the synthetic saturated profiles shown in the right panel of Figure~\ref{synthetic}, both plotted against Tycho--2 BT magnitudes. The vertical axis has an arbitrary offset (the magnitude zero point) which has been adjusted to align the data near BT$\sim$12.  For comparison a 1:1 line is also shown (solid line with $+$ symbols.)  An aperture of 20 pixels ($\sim$ 105 arc sec) was used for both the IRAF measurements and the synthetic photometry.}
\label{derived_mags}
\end{figure}

The mean plate magnitudes for 19 stars, as measured on 90 plates, are shown as open circles in Fig.~\ref{derived_mags} plotted against their BT (Tycho-2) magnitudes.  For comparison a 1:1 relation is shown as the solid line with upright crosses.  The observed relationship between plate and catalogued magnitudes is reasonably linear for the fainter stars, but clearly falls off (by about 1~mag) for the brightest stars as saturation effects become more significant.  That saturation is the cause is clear from the aperture photometry magnitudes determined for the saturated, synthetic profiles shown in the right hand panel of Fig.~\ref{synthetic}, using a circular aperture of 20 pixels.  (The same flux saturation level was used as that which produced the results of Fig.~\ref{radius_comp}.  The vertical offset is arbitrary for the plate and synthetic magnitudes, and has been adjusted so the data overlap near BT$\sim$12, which is the same zero--point as was used to plot the solid line in Fig.~\ref{radius_comp}.)  The synthetic photometry results are shown in Fig.~\ref{derived_mags} by the solid line with no symbols.  The general trend of this line matches quite well the observed plate magnitudes (circles).  It is relatively straightforward to derive a polynomial transformation from plate to standard magnitude$^{6}$. 

\vspace{0.35 cm}
{\it Conclusion}\\

We have not attempted to optimise the fit to the radii of saturation (Fig.~\ref{radius_comp}) by a more careful selection of the synthetic saturation level, or to improve the agreement by seeking a better match between the synthetic profile and the stellar profiles obtained from the Bamberg plates. The purpose of the study was to illustrate that the known, extended stellar radial profile, when digitised as a photo-positive from a photographic plate, can be used successfully for aperture photometry, even though (indeed, {\it because}) the stellar image is mostly saturated.  Such a result was well known to iris photometrists (e.g. Moffat$^{\rm {4}}$, Schaefer$^{\rm {5}}$), but may be less widely appreciated in the modern digital age.

\vspace{0.3 cm}
{\it Acknowledgements} \\

We thank the staff of the Dr. Remeis Observatory, Bamberg, for access to their plate collection for hospitality over several visits. M. Tsvetkov and A. Borisova were supported by the Alexander von Humboldt Foundation under the `Pact of stability of South-East Europe' programme, and grants from BAS/DFG 436-BUL110/120/0-2 and the Bulgarian National Science Fund (NFS I-1103/2001).  D. Coates thanks Prof G. Simon for access to the facilities of the School of Physics and Materials Engineering.  We used IRAF, from the US National Optical Astronomical Observatories, for the photometric measurements. This research has made use of the on--line SIMBAD data facility of the Stellar Data Centre (CDS), Strasbourg, the NASA ADS database, and the Sofia Wide Field Plate Database (WFPDB).  

\vspace{.5cm}

{\it References} \vspace{0.1 cm}\\
(1) King, I.R., 1971, PASP, {\bf 83}, 199. \\
(2) Kormendy, J., 1973, AJ, {\bf 78}, 255. \\
(3) Racine, R., 1996, PASP, {\bf 108}, 699. \\
(4) Moffat, A.F.J., 1969, A\&A, {\bf 3}, 455. \\
(5) Schaefer, B.E., 1981, PASP, {\bf 93}, 253. \\
(6) Innis, J.L., Borisova, A.P., Coates, D.W., Tsvetkov, M.K., 2002, MNRAS, {\bf 355}, 591. \\
(7) de Vaucoulers, G., 1958, Ap. J., {\bf 128}, 465. \\
(8) Strohmeier, W., Mauder, H., 1969, Sky Telesc. {\bf 37}, 10. \\

\end{document}